\documentclass[journal]{IEEEtran}
\usepackage{diagbox}
\usepackage[inline]{enumitem}
\usepackage{xcolor}
\usepackage{amsmath,amsfonts}
\usepackage{algorithm}
\usepackage[group-separator={,},group-minimum-digits=4]{siunitx}
\usepackage{threeparttable}
\usepackage{array}
\usepackage[caption=false,font=normalsize,labelfont=sf,textfont=sf]{subfig}
\usepackage{textcomp}
\usepackage{stfloats}
\usepackage{url}
\usepackage{verbatim}
\usepackage{graphicx}
\usepackage{algpseudocode}
\usepackage{cite}
\usepackage{balance}
\usepackage{mathrsfs}
\usepackage[colorlinks=true,
            linkcolor=blue,
            anchorcolor=blue,
            citecolor=blue]{hyperref}

\begin{document}

\title{
    \fontsize{21 pt}{\baselineskip}\selectfont{Integrated Sensing and Communication Under DISCO Physical-Layer Jamming Attacks}
} 
\author{ 
{       Huan~Huang,~\textit{Member,~IEEE}, 
        Hongliang~Zhang,~\textit{Member,~IEEE},
        Weidong~Mei,~\textit{Member,~IEEE},
        Jun~Li, 
        Yi~Cai,~\textit{Senior~Member,~IEEE}, 
        A.~Lee~Swindlehurst,~\textit{Fellow,~IEEE},
        and~Zhu~Han~\textit{Fellow,~IEEE}
}
\thanks{

H.~Huang, J.~Li, and Y.~Cai are with the School of Electronic and Information Engineering, Soochow University, Suzhou, Jiangsu 215006, China 
(e-mail: hhuang1799@gmail.com, ljun@suda.edu.cn, yicai@ieee.org).

H.~Zhang is  with the 
School of Electronics, Peking University, Beijing 100871, China (email: hongliang.zhang92@gmail.com). 

W.~Mei is with the National Key Laboratory of Wireless Communications,
University of Electronic Science and Technology of China (e-mail: wmei@uestc.edu.cn).



A.~L.~Swindlehurst is with the Center for Pervasive Communications and Computing, University of California, Irvine, CA 92697, USA (e-mail: swindle@uci.edu).

Z.~Han is with the University of Houston, Houston, TX 77004, USA (e-mail: hanzhu22@gmail.com).
}
}
\maketitle

\begin{abstract}
    Integrated sensing and communication (ISAC) systems traditionally presuppose that sensing and communication (S\&C) channels 
    remain approximately constant during their coherence time.
    However, a ``DISCO'' reconfigurable intelligent surface (DRIS), i.e., 
    an illegitimate RIS with random, time-varying reflection properties that acts like a ``disco ball,'' 
    introduces a paradigm shift that enables active channel aging more rapidly during the channel coherence time.
    In this letter, we investigate the impact of DISCO jamming attacks launched by a DRIS-based fully-passive jammer (FPJ) on an ISAC system.
    Specifically, an ISAC problem formulation and a corresponding waveform optimization are presented 
    in which the ISAC waveform design considers the trade-off between the S\&C performance 
    and is formulated as a Pareto optimization problem.
    Moreover, a theoretical analysis is conducted to quantify the impact of DISCO jamming attacks.
    Numerical results are presented to evaluate the S\&C performance under DISCO jamming attacks
    and to validate the derived theoretical analysis.
\end{abstract}


\section{Introduction}\label{Intro}
Integrated sensing and communication (ISAC) is a promising candidate technology for future sixth generation (6G) wireless communication systems and has attracted increasing attention. 
In particular, ISAC implements joint target sensing and data communication 
using the same RF hardware and computing platform~\cite{HLZhangISAC,HeathISAC}, 
offering an exciting opportunity to implement sensing using traditional wireless communication infrastructure~\cite{LiuJSAC}.
The added sensing functionality enabled by the collection of environmental data makes ISAC a fundamental component of future smart environments. 
In particular, ISAC is applicable to vehicle-to-everything communications, smart homes, and smart manufacturing. 
There has been considerable research on ISAC-related problems, including ISAC waveform design~\cite{FLiuISAC,FLiuISAC1}.
Considering the difficulties of optimization problems involving the mean squared error (MSE) or the Cram$\acute{\rm e}$r-Rao lower bound (CRLB), 
most existing works use simpler alternative optimization criteria such as the transmit beampattern~\cite{LSWCMISAC,FLiuISAC,FLiuISAC1}.

Recently, reconfigurable intelligent surfaces (RISs) are also anticipated to play a critical role in future 6G wireless communications. 
These surfaces are embedded with many elements 
whose reflection coefficients can be tuned by simple programmable PIN or varactor diodes~\cite{IRSCuiTJ,IRSsur1,IRSsur2,ZhangPost}.
By properly tuning these reflection coefficients, 
the electromagnetic environment can be reshaped to enhance signal transmission to improve both sensing and communication (S\&C)~\cite{LSWCMISAC,STARISAC,ARIISISAC}.
However, the S\&C performance enhancement relies on the \emph{basic premise} that the S\&C channels are approximately constant during the channel coherence time.
This basic premise is normally valid, but can be negated when so-called DISCO RISs (DRISs) are deployed~\cite{MyWCMag}.

The DRIS concept was first introduced with DRIS-based fully-passive jammer (FPJ)~\cite{MyTVT}, 
where a DRIS with time-varying reflection properties acts like a ``disco ball.''
The DRIS introduces active channel aging (ACA), and 
thus the wireless channels will vary more rapidly than the channel coherence time~\cite{MyTVT,DIRSTWC,TWCAnti}.
Such ACA can be used to jam communication users or degrade the accuracy of target sensing 
without the use of either jamming power or channel state information (CSI).
This type of ACA interference (ACAI) is referred to as a DISCO jamming attack, 
and renders the above basic premise for ISAC systems invalid.

In this work, we aim to characterize the impact of DISCO jamming attacks on ISAC systems. 
To the best of our knowledge, this is the first time that the validity of the basic premise has been investigated for ISAC systems.
The main contributions are summarized as follows:
\begin{itemize}
\item 
A DRIS-based FPJ is introduced into an ISAC system to launch DISCO jamming attacks. 
Furthermore, a practical RIS model is considered for the DRIS-based FPJ, where the DRIS phase shifts of the reflective elements are discrete and
the DRIS amplitudes are a function of their corresponding phase shifts.  
\item  
In the ISAC waveform design, the desired sensing waveform and the signal-to-interference-plus-noise ratio (SINR) 
are used as S\&C performance metrics. 
Consequently, the ISAC waveform design problem under DISCO jamming attacks is formulated as a Pareto optimization problem.
We present a corresponding ISAC waveform design 
by considering the trade-off between the S\&C performance metrics.
\item  
We show that the DISCO jamming attacks 
lead to biased estimation of the target parameters
and impair the communication sum rate.
Moreover, a theoretical analysis is performed to quantify the impact of the DISCO jamming attacks. 
\end{itemize}


\emph{Notation:} We employ bold capital letters for a matrix, e.g., ${\bf{X}}$, lowercase bold letters for a vector, e.g., ${\boldsymbol x}_l$, 
and italic letters for a scalar, e.g., $k$. 
The superscripts $(\cdot)^{T}$ and $(\cdot)^{*}$ represent the transpose and the Hermitian transpose, respectively, 
and the symbols $\|\cdot\|_{\rm F}$ and $|\cdot|$  represent the Frobenius norm and the absolute value, respectively. 

\section{System Description}\label{Princ}
In this section, we first illustrate DISCO jamming attacks launched by a DRIS-based FPJ in Section~\ref{DISCOJamm}.
Then, the model of the ISAC system under DISCO jamming attacks is given in Section~\ref{DIRSFPJComm}.
\begin{figure}[!t]
    \centering
    \includegraphics[scale=0.6]{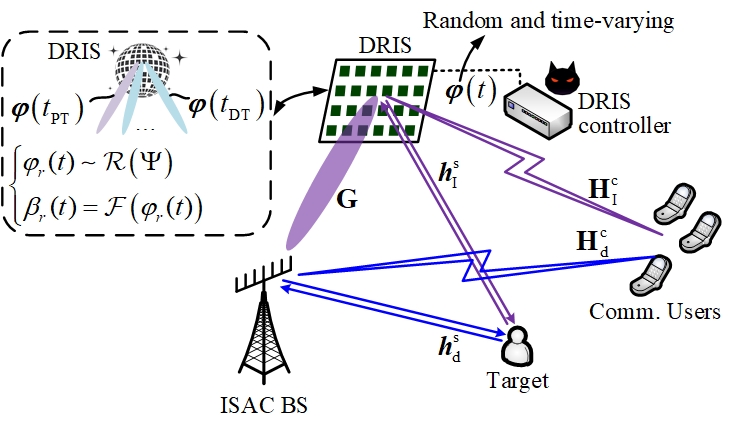}
    \caption{The downlink of an ISAC system jammed by a DRIS-based FPJ,
    where the time-varying DRIS reflection coefficients are randomly generated by the DRIS controller.}
    \label{fig1}
\end{figure}

\subsection{DRIS-Based Active Channel Aging}\label{DISCOJamm}
Fig.~\ref{fig1} shows an ISAC system under DISCO  jamming attacks 
launched by a DRIS-based FPJ~\cite{MyTVT}.
The DRIS with $N_{\rm D} = N_{{\rm D},h} \times N_{{\rm D},v}$ reflective elements is implemented using PIN diodes, whose ON/OFF behavior
only allows for discrete phase shifts. 
Therefore, the time-varying DRIS phase shifts
${\varphi}_{ r}(t)$ ($r=1,\cdots, N_{\rm \!D})$ are randomly selected from 
a discrete set $\Psi$ with $b$-bit quantized values
$\left\{ {\phi_{1}},\cdots,  {\phi_{2^{b}}} \right\}$
and follow a stochastic distribution denoted as ${\varphi}_{ r}(t) \sim {\mathcal R}\!\left(\! \Psi\right)$.
The amplitude of the reflection coefficient $\beta_r$ is a function of ${\varphi}_{ r}(t)$
and represented by $\beta_r(t) = {\mathcal F}\left({\varphi}_{ r}(t)\right)$, 
where ${\mathcal F}\left(\Psi\right) = {\Xi} = \left\{\mu_1,\cdots, \mu_{2^b}\right\}$.
As a result, the time-varying DRIS reflecting vector ${\boldsymbol{\varphi} ({t})}$ is given by $\boldsymbol{\varphi}\left( {{t}} \right) = 
    \left[ {{\beta _1}({t}){e^{j{\varphi _1}({t})}}, \cdots ,{\beta _{{N_{\rm{D}}}}}\!({t}){e^{j{\varphi _{{N_{\rm{D}}}}}({t})}}} \right]$.

In traditional wireless systems, the wireless channels are assumed to be fixed  during the channel coherence time.
Consequently, the CSI estimated during the pilot transmission (PT) phase can be used to  design the waveform used in the remaining 
data transmission (DT) phase of  each channel coherence time.
Before S\&C data transmission, the ISAC base station (BS) first learns the CSI during the PT phase via existing methods 
such as the least squares (LS) algorithm.
Mathematically, the CSI$\footnote{We assume that perfect CSI is available
as imperfect CSI is not a primary concern in the jamming scenario, and its impact has been thoroughly studied~\cite{RefSLNRadd}.}$ 
estimated during the PT phase is written as
\begin{equation}
    {{\bf{H}}_{{\rm{\!P\!T}}}^{\rm c}} =  {{\bf{H}}_{\rm{d}}^{\rm c} \!+\! {{\bf{G}} }{ \rm{diag}\!} \left( {\boldsymbol{\varphi} ({t_{\rm{\!P\!T}}})} \right){\! \bf{H}}_{\rm{I}}^{\rm c} } ,
    \label{PTCSI}
\end{equation}
where ${\bf{H}}_{\rm{d}}^{\rm c}$ and ${\bf{H}}_{\rm{D}}^{\rm c}  = {{\bf{G}}}{ \rm{diag}\!} \left( {\boldsymbol{\varphi} ({{t_{\rm{\!P\!T}}}})} \right) {\bf{H}}_{\rm{I}}^{\rm c}$
represent the direct channel and 
the time-varying DRIS-jammed channel between the ISAC BS and the communication users, respectively.

Due to the random and time-varying DRIS reflecting coefficients, ACA is introduced within the channel coherence time.
The time between changes in the DRIS reflection coefficients is typically assumed to be about
the same as the length of the PT phase~\cite{MyWCMag,TWCAnti}. 
As a result, the DRIS rapidly ages the wireless channels, and effectively produces a
channel with a coherence interval approximately equal to the PT phase.
Mathematically, the ACA channel during the DT phase can be represented as
\begin{equation}
    {{\bf{H}}^{\rm c}_{{\rm{\!A\!C\!A}}}}  \!=\! {{\bf{H}}^{\rm c}_{{\rm{\!D\!T}}}}\!-\!{{\bf{H}}^{\rm c}_{{\rm{\!P\!T}}}}  
    \!=\!{{\bf{G}}}{ \rm{diag}\!} \left( {\boldsymbol{\varphi} ({t_{\rm{\!D\!T}}})} \!-\!{\boldsymbol{\varphi} ({t_{\rm{\!P\!T}}})} \right){\! \bf{H}}_{\rm{I}}^{\rm c},
    \label{ACAEq}
\end{equation}
where  $\bf G$ and ${\bf H}_{\rm I}^{\rm c}$ denote the channel between the ISAC BS and the DRIS
and the channel between the DRIS and all communication users.

\subsection{ISAC Under DISCO Jamming}\label{DIRSFPJComm}
\underline{\textit{Communication Model:}}
In Fig.~\ref{fig1}, the ISAC BS is equipped with $N$ transmit antennas
to communicate with $K_{\rm c}$ single-antenna communication users.
During the DT phase, the ISAC BS transmits $L$ symbols to these users,
and thus the length of the data frame is $L$.
Consequently, the received signals at the users are expressed as
\begin{equation}
    {\bf{Y}\!_{\rm c}} = {\bf{S}} + \underbrace {\left( {{{\bf{H}}^{\rm c}_{{\rm{PT}}}}{\bf{X}} - {\bf{S}}} \right)}_{\rm{MUI}} + 
    \underbrace {{{\bf{H}}^{\rm c}_{{\rm{\!A\!C\!A}}}} {\bf{X}}}_{\rm{ACAI}} + {\bf{N}_{\rm c}},
    \label{SigRx}
\end{equation}
where $\bf{S}$ denotes the $K \times L$ desired constellation symbol matrix,
${\bf X}=[{\boldsymbol x}^T_1,\cdots, {\boldsymbol x}^T_L ]^T$ represents the $N \times L$ transmitted signal matrix used as the ISAC waveform
for both the communication and sensing functions~\cite{FLiuISAC,FLiuISAC1},
and ${\bf N}_{\rm c}$ is a $K \times L$ Gaussian noise matrix composed of independent and identically distributed (i.i.d.) elements 
with zero mean and variance $\sigma_{\rm c}^2$, i.e., ${{n}^{\rm c}_{k,l}} \sim \mathcal{CN}\left(0,\sigma_{\rm c}^2\right)$.
Based on~\eqref{SigRx}, we can see that the DRIS-based FPJ imposes ACAI
on the signals in addition to multi-user interference (MUI)~\cite{FLiuISAC}.
Referring to~\cite{DIRSTWC,FLiuISAC}, the SINR
at the $k$-th user is given by
\begin{equation}
    {\gamma _k} \!=\! \frac{{\mathbb{E}\!\left[ \!{{{\left| {{s_{k,l}}} \right|}^2}} \right]}}{{{\mathbb E}\!\left[\! \left|{\left(\! {{{ ( {\boldsymbol{h}_{{\rm{PT}},k}^{\rm{c}}} )}^*}{\boldsymbol{x}_l} \!-\! {s_{k,l}}} \right) 
    \!+\! {{ \left( \!{\boldsymbol{h}_{{\rm{A\!C\!A}},k}^{\rm{c}}} \right)}^*}{\boldsymbol{x}_l}}\! \right|^2\right] \!+\! {\sigma_{\rm c}^2}}}.
    \label{SINRDRIS}
\end{equation}
Consequently, the sum rate can be computed based on~\eqref{SINRDRIS}, i.e., $R_{\rm{sum}} = \sum\nolimits_{k = 1}^K {{R_k}}  = \sum\nolimits_{k = 1}^K {{{\log }_2}\left( {1 + {\gamma _k}} \right)}$.

\underline{\textit{Sensing Model:}}
To quantify the sensing performance,
we use the quality of the estimated target angle $\theta$~\cite{NearfieldMo1}.
More specifically, the $L$ symbols are reflected by the target and then received at the ISAC BS, resulting in~\cite{STARISAC}
\begin{equation}
    {{\bf{Y}}_{\rm{\! s}}} = \chi\!\left( {\boldsymbol{h}_{\rm d}^{\rm s}} + {\boldsymbol{h}_{\rm D}^{\rm s}}\right) \!
    \left( {\boldsymbol{h}_{\rm d}^{\rm s}} + {\boldsymbol{h}_{\rm D}^{\rm s}}\right)^{*}\!
    {\bf X} + {\bf N}_{\rm s},
    \label{SensingChannel}
\end{equation}
where $0\le \!\chi \!\le 1$ represents the reflection cross-section of the target, 
${\boldsymbol{h}_{\rm d}^{\rm s}} = {\sqrt{{{\mathscr{L}}^{\rm s}_{\!{\rm {d}}}}}}{\boldsymbol{\alpha}}\!\left(\! {N,\theta }\right)$ denotes the direct sensing path,
${\boldsymbol{h}}_{\rm{D}}^{\rm{s}} = {\sqrt{{{\mathscr{L}}^{\rm s}_{\!{\rm {cas}}}}}}{\bf G}{\rm diag}\left(\! {\boldsymbol{\varphi}}(t)\right)\!{\boldsymbol{\alpha}}\!\left(\! {{N_h},{N_v},{\phi _h},{\phi _v}} \right)$ denotes the time-varying DRIS-jammed sensing path,
 ${{\mathscr{L}} _{{\rm {d}}}^{\rm s}}$ and ${{\mathscr{L}}_{{\rm {cas}}}^{\rm s}}$ are the large-scale channel fading coefficients of ${\boldsymbol{h}_{\rm d}^{\rm s}}$ and ${\boldsymbol{h}_{\rm D}^{\rm s}}$,
and ${\bf N}_{\rm s}$ is an $N \times L$ noise matrix whose i.i.d. elements have zero mean and variance $\sigma_{\rm s}^2$.
Furthermore, the steering vectors of the ISAC BS antenna array and the DRIS are respectively defined as
\begin{equation}
    {\boldsymbol{\alpha}} \!\left( N,\theta  \right) \!=\!  
    {\left[ \!{1,{e^{j2\pi \Delta \!\sin \theta }}, \cdots ,{e^{j2\pi (N \!-\! 1)\Delta \!\sin \theta }}} \right]^T}
    \label{LPA}
\end{equation}
and
\begin{equation}
    {\boldsymbol{\alpha}}\!\left( {{N_h},{N_v},{\phi _h},{\phi _v}} \right) \!= \! {\boldsymbol{\alpha}} \!\left({N_h},{\phi _h}\right)\! 
    \otimes {\boldsymbol{\alpha}} \!\left({N_v},{\phi _v}\right),
    \label{UPA}
\end{equation}
where $\Delta$ denotes the array spacing normalized by the wavelength,
and $\otimes $ represents the Kronecker product.

To estimate $\theta$, we exploit the MUSIC algorithm~\cite{DectP}.
The sample covariance matrix computed based on $L$ snapshots in~\eqref{SensingChannel} is 
\begin{equation}
    {\widetilde {\bf{R}}_{\bf{X}}} = \frac{1}{L}{{\bf{Y}}_{\rm{s}}}{\bf{Y}}_{\rm{s}}^*,
    \label{MUSIC_A}
\end{equation}
and the MUSIC spectral function $V\!(\vartheta) $ is then computed from ${\widetilde {\bf{R}}_{\bf{X}}}$.
However, based on~\eqref{SensingChannel}, the time-varying DRIS-jammed sensing path ${\boldsymbol{h}_{\rm D}^{\rm s}}$ 
has been introduced into ${\widetilde {\bf{R}}_{\bf{X}}}$,
which perturbs the location of the spectral peak in $V\!(\vartheta)$ and leads to a biased DoA estimation.

\section{ISAC Waveform Design Under DISCO Jamming}\label{ISACDesign}
In this section, we first formulate the ISAC waveform design problem under DISCO jamming attacks 
and give a waveform optimization design for the ISAC system under these attacks in Section~\ref{ProblemFor}.
In Section~\ref{DRISJamm}, a theoretical analysis is derived to quantify the impact of the DISCO jamming attacks.
\subsection{Problem Formulation And ISAC Waveform Design}~\label{ProblemFor}
According to~\eqref{SigRx} and~\eqref{SINRDRIS}, the optimum ISAC waveform should be designed
to minimize the power of the multi-user interference (MUI) and ACAI.
However, due to the time-varying and random DRIS reflecting vector ${\boldsymbol{\varphi} ({t})}$,
the ISAC BS can not obtain ${\bf H}^{\rm c}_{\rm{A\!C\!A}}$. 
Therefore, we consider designing the ISAC waveform by minimizing the MUI. 
Mathematically, the ISAC waveform design problem is formulated as
\begin{alignat}{1}
    \left( {{\rm{P}}1} \right): &\mathop {\min}\limits_{{\bf{X}}} \left\| {{\bf{H}}_{{\rm{PT}}}^{\rm{c}}{\bf{X}} - {\bf{S}}} \right\|_{\rm{F}}^2 \label{P1forS}\\
    &{\rm{s}}.{\rm{t}}.~{{\bf{C}}_{\bf{X}}} = \frac{1}{L}{\bf{X}}{{\bf{X}}^*} = \frac{{{P_0}}}{N}{{\bf{I}}_N},
    \label{ISACWaveLim}
\end{alignat}
where $P_0$ is the total transmit power at the ISAC BS and
${\bf I}_N$ is an $N \times N$ identity matrix. We assume $N \le L$ to ensure that ${{\bf{C}}_{\bf{X}}}$ is positive-definite.

The strict equality constraint~\eqref{ISACWaveLim} ensures that
the ISAC waveform has the same properties as the best sensing waveform, although the communication performance may be degraded as a result~\cite{FLiuISAC}. 
Therefore, there should be a trade-off between the sensing capability and the communication rate in $\left( {{\rm{P}}1} \right)$,
so we introduce a trade-off factor $\kappa$ ($0 \le \kappa \le 1$) into $\left( {{\rm{P}}1} \right)$,
Denoting the solution to $\left( {{\rm{P}}1} \right)$ as ${\bf X}_0$, the following Pareto optimization problem can be obtained:
\begin{alignat}{1}
    \left( {{\rm{P}}2} \right): &\mathop {\min}\limits_{{\bf{X}}} \kappa \left\| {{\bf{H}}_{{\rm{PT}}}^{\rm{c}}{\bf{X}} - {\bf{S}}} \right\|_{\rm{F}}^2 + (1-\kappa)\left\| {{\bf{X}} - {{\bf{X}}_0}} \right\|_{\rm{F}}^2 \\
    &\,{\rm{s}}.{\rm{t}}.\;{\left\| {\bf{X}} \right\|_{\rm{F}}^2} = LP_0.
    \label{P1forISA}
\end{alignat}
For different $\kappa$, the solution to $\left( {{\rm{P}}2} \right)$ makes different trade-offs between S\&C performance.
More specifically, the smaller the $\kappa$, the better the sensing performance, but the worse the communication performance.
 
To solve $\left( {{\rm{P}}2} \right)$, we first compute
${\bf X}_0$ from $\left( {{\rm{P}}1} \right)$. 
Since $\left( {{\rm{P}}1} \right)$
is a classical orthogonal Procrustes problem, a simple closed-form solution to $\left( {{\rm{P}}1} \right)$ 
can be obtained based on the singular value decomposition (SVD), i.e.,
${{\bf{X}}_0} = \sqrt {\frac{{{P_0}L}}{N}} {\bf{U}}{{\bf{I}}_{N \times L}}{\!{\bf{V}}^*}$,
where $\bf U$ and $\bf V$ are the left and right singular value matrices of ${\bf H}^{\rm c}_{\rm{P\!T}}{\bf S}$ 
and satisfy ${\bf U} {\boldsymbol \Sigma} {\bf V}^* = {\bf H}^{\rm c}_{\rm{P\!T}}{\bf S}$.

Consequently, we transform Problem $\left( {{\rm{P}}2} \right)$ into the following form using the approach of~\cite{FLiuISAC},
\begin{alignat}{1}
    \left( {{\rm{P}}2}-{{\rm{E}}1} \right): &\mathop {\min}\limits_{{\bf{X}}} \left\| {{\bf{AX}} - {\bf{B}}} \right\|_{\rm{F}}^2 \\
    \nonumber
    &\,{\rm{s}}.{\rm{t}}.~\eqref{P1forISA},
\end{alignat}
where ${\bf A} = { [ {\sqrt \kappa  {{({\bf{H}}_{{\rm{PT}}}^{\rm{c}})}^T},\sqrt {1 - \kappa } {{\bf{I}}_N}} t]^T}$ 
and ${\bf{B}} = { [ {\sqrt \kappa  {{\bf{S}}^T},\sqrt {1 - \kappa } {\bf{X}}_0^T}]^T}$.
 It is worth noting that $\left( {{\rm{P}}2}-{{\rm{E}}1} \right)$ can be transformed into a semidefinite programming (SDP) problem 
using semidefinite relaxation (SDR),
and it has only one quadratic constraint, i.e.,~\eqref{P1forISA}.
Therefore, the rank-1 SDR solution is its globally optimal solution.

\subsection{Impact of DISCO Jamming Attacks}~\label{DRISJamm}
For a given ISAC waveform, the sum rate is affected not only by the MUI due to sensing functionality considerations, but also by the ACAI.
The impact of MUI has been investigated in the
existing literature, such as~\cite{FLiuISAC,FLiuISAC1} so we focus on the impact of the ACAI imposed by the  DRIS-based FPJ.
For ease of presentation, we rewrite the interference term in~\eqref{SINRDRIS} as 
\begin{alignat}{1}
    \nonumber
    {\mathscr{I}} &= \! \left| {{{ ( {\boldsymbol{h}_{{\rm{PT}},k}^{\rm{c}}} )}^*}{\boldsymbol{x}_l} \!-\! {s_{k,l}}} \right|^2 \!\!+\! 
    \left({{{ ( {\boldsymbol{h}_{{\rm{PT}},k}^{\rm{c}}} )}^*}{\boldsymbol{x}_l} \!-\! {s_{k,l}}} \right){\boldsymbol{x}^*_l}{\boldsymbol{h}_{{\rm{ACA}},k}^{\rm{c}}} \\
    &+  \!\!({{{\! (\! {\boldsymbol{h}_{{\rm{PT}},k}^{\rm{c}}} )}^*}{\!\boldsymbol{x}_l} \!-\! {s_{k,l}}}  )^*{{ \!( {\boldsymbol{h}_{{\rm{ACA}},k}^{\rm{c}}} )}^*}\!{\boldsymbol{x}_l} \!+ \!\left|{{ ( {\boldsymbol{h}_{{\rm{ACA}},k}^{\rm{c}}} )}^*}{\boldsymbol{x}_l} \right|^2. 
    \label{RewInterf}
\end{alignat}

The DRIS must be equipped with a large number of reflective elements to overcome the multiplicative propagation loss 
in the DRIS-jammed channels.
The elements of ${\bf{H}}^{\rm c}_{\rm{A\!C\!A}}$ have the statistical characteristics outlined
in Proposition~\ref{Proposition1}.
\newtheorem{proposition}{Proposition}
\begin{proposition}
    \label{Proposition1}
    The elements of ${\bf{H}}^{\rm c}_{\rm {\!A\!C\!A}}$ converge in distribution to $\mathcal{CN}\!\left( {0,  {{{\mathscr{L}}\!_{{\rm {cas}},k}} {N\!_{\rm D}} {\overline \mu} } } \right)$ as $N_{\rm D} \to \infty$, i.e.,
    \begin{equation}
        {\left[ {{\bf H}^{\rm c}_{\rm {\!A\!C\!A}}} \right]_{n,k}} \mathop  \to \limits^{\rm{d}} \mathcal{CN}\!\left( {0,  {{{\mathscr{L}}^{\rm c}_{{\rm {cas}},k}} {N\!_{\rm D}}{\overline \mu} } } \right), \forall n,k,
        \label{HDSta}
    \end{equation}
where ${{\mathscr{L}}^{\rm c}_{{\rm {cas}},k}}$ is the large-scale channel
fading coefficient of the DRIS-jammed channel between the ISAC BS and the $k$-th communication user,
$\overline \mu = \sum\nolimits_{i1 = 1}^{{2^b}} \sum\nolimits_{i2 = 1}^{{2^b}}{p_{i1}}{p_{i2}} \big( {\mu _{i1}^2} + {\mu _{i2}^2} - 2{\mu _{i1}}{\mu _{i2}}$ $\cos ( {{\phi_{i1}} - {\phi_{i2}}} ) \big)$,
    ${\mu _{i1}}$, ${\mu _{i2}} \in \Xi$, ${\phi_{i1}}$, ${\phi_{i2}} \in \Psi$, and ${p_{i1}}, {p_{i2}}$ 
are the probabilities of the random phases ${\phi_{i1}}, {\phi_{i2}}$.
\end{proposition}

\begin{IEEEproof}
    See~\cite{TWCAnti}.
\end{IEEEproof}

Based on Proposition~\ref{Proposition1}, the impact of the ACAI is mathematically quantified by Theorem~\ref{Theorem1} below. 
\newtheorem{theorem}{Theorem}
\begin{theorem}
\label{Theorem1}
Under DISCO jamming attacks launched by the DRIS-based FPJ, a lower
bound for the SINR received at the $k$-th communication user
${\gamma _k}$ for any ISAC waveform computed from $\left( {{\rm{P}}2}-{{\rm{E}}1} \right)$ converges in distribution to 
\begin{alignat}{1}
    \nonumber
    {\gamma _k} & \ge \frac{{\mathbb{E}\!\!\left[\! {{{\left|{{s_{k,l}}} \right|}^2}} \right]}}{{{\mathbb E}\!\!\left[\! \left|\! {\left( {{{ ( {\boldsymbol{h}_{{\rm{PT}},k}^{\rm{c}}} )}^*}{\boldsymbol{x}_l} - {s_{k,l}}} \right) }\!  \right|^2\right] \!\!+\!{\mathbb E}\!\!\left[\!\left\|\! {\boldsymbol{h}_{{\rm{A\!C\!A}},k}^{\rm{c}}} \!\right\|^2 \!\right]\!{\mathbb E}\!\!\left[\!\left\|{\boldsymbol{x}_{l}} \!\right\|^2\right]  \!+\!{\sigma^2}}}, \\
    &\mathop  \to \limits^{\rm{d}} \!\frac{{\mathbb{E}\!\!\left[ \!{{{\left|{{s_{k,l}}} \right|}^2}} \right]}}{{{\mathbb E}\!\!\left[ \!\left|\! {\left( {{{ ( {\boldsymbol{h}_{{\rm{PT}},k}^{\rm{c}}} )}^*}{\boldsymbol{x}_l} - {s_{k,l}}} \right) }\!  \right|^2\right] +{P_0 {{{\mathscr{L}}^{\rm c}_{{\rm {cas}},k}} {N\!_{\rm D}}{\overline \mu} }} +{\sigma^2}}}.
    \label{SINRDRISMath}
\end{alignat}
\end{theorem}

\begin{IEEEproof}
Since the ISAC waveform $\bf X$ computed from $\left( {{\rm{P}}2}-{{\rm{E}}1} \right)$ is optimized only based on ${\bf H}^{\rm c}_{\rm{P\!T}}$,
we can assume that $\bf X$ is independent of the ACA communication channel ${\bf H}^{\rm c}_{\rm{A\!C\!A}}(t)$.
Consequently, the expectation of ${\mathscr{I}}$ in~\eqref{RewInterf} is
\begin{equation}
    {\mathbb E}\! \left[ {\mathscr{I}}\right] \!=\! {\mathbb E}\!\left[\left| {{{ ( {\boldsymbol{h}_{{\rm{P\!T}},k}^{\rm{c}}} )}^*}{\boldsymbol{x}_l} \!-\! {s_{k,l}}}\! \right|^2\right] 
    \!+ {\mathbb E}\!\left[\left| {{ \left( \!{\boldsymbol{h}_{{\rm{A\!C\!A}},k}^{\rm{c}}} \!\right)\!}^*}\!{\boldsymbol{x}_l}\!\right|^2\right] .
    \label{InterferTermReduce}
\end{equation}

According to the Cauchy--Schwarz inequality and Proposition~\ref{Proposition1}, we have 
\begin{equation}
    {\mathbb E}\!\left[\left| {{ \left( \!{\boldsymbol{h}_{{\rm{A\!C\!A}},k}^{\rm{c}}}\!\right)\!}^*}\!{\boldsymbol{x}_l}\!\right|^2\right]
    \le {\mathbb E}\!\!\left[\!\left\|\! {\boldsymbol{h}_{{\rm{A\!C\!A}},k}^{\rm{c}}} \!\right\|^2 \!\right]\! {\mathbb E}\!\!\left[\!\left\|{\boldsymbol{x}_{l}} \!\right\|^2\right]
    {\mathop \to \limits^{\rm{d}}}  P_0 {{{\mathscr{L}}^{\rm c}_{{\rm {cas}},k}} {N\!_{\rm D}}{\overline \mu} }.
    \label{InterferTermReduceFirst}
\end{equation}
Substituting~\eqref{InterferTermReduce} and~\eqref{InterferTermReduceFirst} into~\eqref{SINRDRIS}, \eqref{SINRDRISMath} is derived.
\end{IEEEproof}

The numerical results given in the next section will indicate that 
the waveform design based on $\left( {{\rm{P}}2}-{{\rm{E}}1} \right)$ achieves performance that is very close to the derived lower bound in Theorem~\ref{Theorem1}.
This indicates that without any knowledge of the DRIS-jammed channels, 
there is no ``better'' waveform design that could provide a significant improvement over the lower bound.
Fortunately, Proposition~\ref{Proposition1} gives the statistical characteristics of the ACA channel ${\bf H}_{\rm{ACA}}$,
which may provide a potential approach to designing an anti-jamming scheme~\cite{TWCAnti}.

\section{Simulation Results and Discussion}\label{ResDis}
We consider an ISAC system equipped with a 16-element antenna array located at (0m, 0m, 3m) and jammed by the DRIS-based FPJ.
The ISAC BS communicates with 8 single-antenna users 
that are randomly distributed in the circular region $S$ centered at (0m, 180m, 0m) with a radius of 20m. 
The DRIS with 1024 ($N_{{\rm D},h} = 32,N_{{\rm D},v} = 32 $) reflective elements is deployed at (2m, 0m, 2m) to launch DISCO physical-layer jamming attacks.
We assume that the DRIS has one-bit quantized phase shifts and gain values taken from $\Psi = \{\frac{\pi}{9},\frac{7\pi}{6}\}$ and $\Xi = {\cal F}\!\left({\Theta}\right) = \{0.8,1\}$~\cite{IRSsur1},
and the two phase shifts are chosen with equal probability. Consequently, $\overline \mu$ in Theorem~\ref{Theorem1} is 1.6078.
The length of the data frame is $L = 18$ 
and the trade-off factor in~$\left( {{\rm{P}}2} \right)$ is $\kappa = 0.2$.

Based on the settings above, the wireless channel $\bf G$ is constructed using a
near field channel model, 
while the wireless channels ${\bf H}_{\rm I}^{\rm c}$ and ${\bf H}_{\rm d}^{\rm c}$ 
are both based on far field channel models~\cite{TWCAnti,DIRSTWC,MyTVT,NearfieldMo1}.
The large-scale line-of-sight (LoS) and non-line-of-sight (NLoS) channel fading coefficients  are defined in Table~\ref{tab1} based on 3GPP propagation models~\cite{3GPP}, 
and the variance of the noise is $\sigma^2_{\rm c}\!=\!-170\!+\!10\log_{10}\left(BW\right)$ dBm with a transmission bandwidth of 180 KHz.
\begin{table}
    \footnotesize
    \centering
    \caption{Wireless Channel Simulation Parameters}
    \label{tab1}
    \begin{threeparttable}
    \begin{tabular}{ c|c }
    \hline
    Parameter        &Value\\
    \hline
    Large-scale LoS fading       & $ 35.6 + 22{\log _{10}}({d}) $ (dB) \\
    \hline
    Large-scale NLoS fading      &$32.6+36.7{\log _{10}}({d})$ \\
    \hline
    \end{tabular}
    \end{threeparttable}
\end{table} 

\begin{figure}[!t]
    \centering
    \includegraphics[scale=0.59]{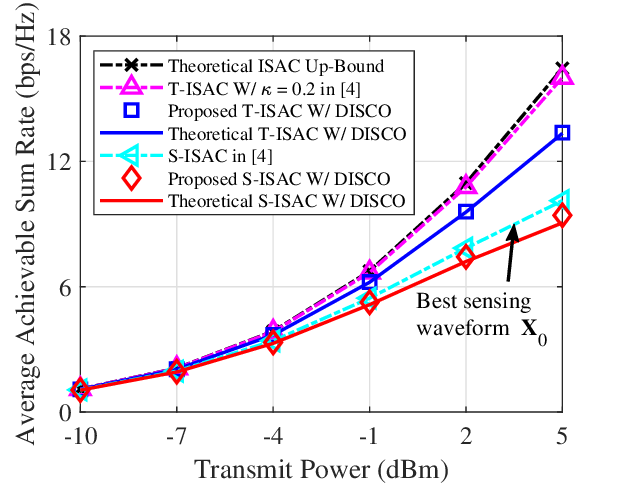}
    \caption{Average achievable sum rate vs. total power.}
    \label{fig3}
\end{figure}

Fig.~\ref{fig3} illustrates the results obtained by the following approaches:
1) the sum rate obtained without MUI or ACAI (Theoretical ISAC Up-Bound);
2) the sum rate achieved by a traditional ISAC system~\cite{FLiuISAC} (T-ISAC W/ $\kappa = 0.2$ in~\cite{FLiuISAC});
3) the sum rate achieved by an ISAC system under DISCO jamming attacks, i.e., $\left( {{\rm{P}}2}-{{\rm{E}}1} \right)$ (Proposed T-ISAC W/ DISCO);
4) the theoretical analysis of Proposed T-ISAC W/ DISCO based on Theorem~\ref{Theorem1} (Theoretical T-ISAC W/ DISCO);
5) the sum rate achieved by the ISAC system~\cite{FLiuISAC} with the strict equality constraint~\eqref{ISACWaveLim} (S-ISAC in~\cite{FLiuISAC});
6) the sum rate achieved by an ISAC system under DISCO jamming attacks with the strict equality constraint~\eqref{ISACWaveLim},
i.e., $\left( {{\rm{P}}1}\right)$ (Proposed S-ISAC W/ DISCO);
7) the theoretical analysis of Proposed S-ISAC W/ DISCO based on Theorem~\ref{Theorem1} (Theoretical S-ISAC W/ DISCO).

We can see from Fig.~\ref{fig3} that by considering the trade-off between the S\&C performance,
the gap between Theoretical ISAC Up-Bound and T-ISAC W/ $\kappa = 0.2$ in~\cite{FLiuISAC} is small.
In other words, the sum rate is not seriously affected by the sensing functionality when the ISAC waveform is well designed.
However, the performance is severely compromised by DISCO jamming attacks.
Without any knowledge of the DRIS-jammed channels, the results of Proposed T-ISAC W/ DISCO and Proposed S-ISAC W/ DISCO are very close to 
the ideal lower-bounded performance of Theoretical T-ISAC W/ DISCO and Theoretical S-ISAC W/ DISCO.
Therefore, the use of the statistical characteristics in~Proposition~\ref{Proposition1}
as side knowledge to improve the ISAC waveform design in Section~\ref{ISACDesign} is a worthwhile pursuit.

\begin{figure}[!t]
    \centering
    \includegraphics[scale=0.58]{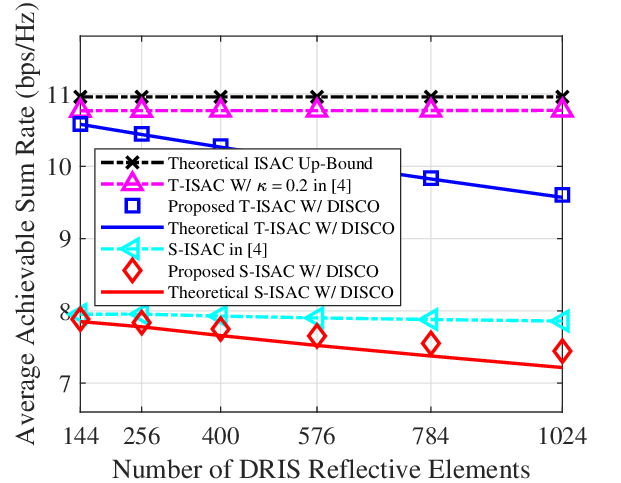}
    \caption{Average achievable sum rate vs. the number of DRIS relfective elements, where the transmit power is 2 dBm.}
    \label{fig4}
\end{figure}

Based on Theorem~\ref{Theorem1}, the impact of DISCO jamming attacks on the sum rate can be quantified by $P_0{{{\mathscr{L}}^{\rm c}_{{\rm {cas}},k}} {N\!_{\rm D}} {\overline \mu}}$.
To evaluate the validity of Theorem~\ref{Theorem1}, the relationship between the sum rate and the number of DRIS reflective elements 
is given in Fig.~\ref{fig4}.
We can see that the sum rate decreases with the number of DRIS elements $N_D$, 
and the achieved sum rates for different $N_{\rm D}$ are close to the theoretical lower bound.

\begin{figure}[!t]
    \centering
    \includegraphics[scale=0.58]{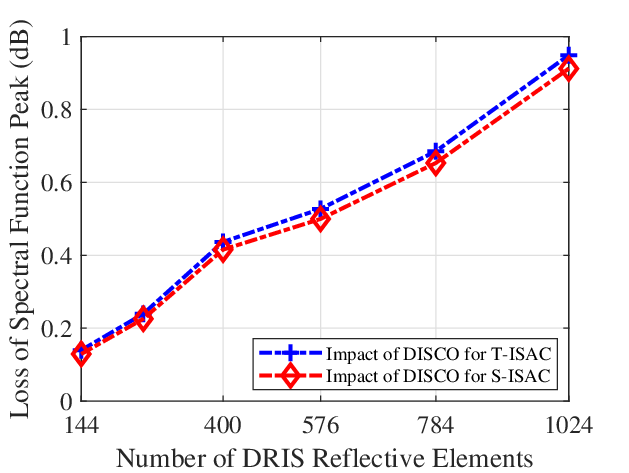}
    \caption{Difference between the spectral function peaks.}
    \label{fig5}
\end{figure}

Fig.~\ref{fig5} shows the impact of the ISAC waveform and the DISCO jamming attacks on the sensing performance.
We assume that the echo SNR from the direct sensing path ${\boldsymbol{h}}^{\rm s}_{\rm d}$ is 10 dB.
We denote the spectral functions obtained from the best sensing waveform ${\bf X}_{0}$ without DISCO jamming attacks and
${\bf X}_{0}$ under DISCO jamming attacks, the ISAC waveform $\bf X$ obtained from $\left( {{\rm{P}}2}-{{\rm{E}}1} \right)$
without DISCO jamming attacks, and $\bf X$ under DISCO jamming attack as $V_{\rm S}(\theta)$, $V_{\rm S-D}(\theta)$, 
$V_{\rm T}(\theta)$, and $V_{\rm T-D}(\theta)$, respectively.
The loss of spectral function peaks~\cite{FLiuISAC1,NearfieldMo1} between $V_{\rm S}(\theta)$ and $V_{\rm S-D}(\theta)$ is 
shown by the red diamond (Impact of DISCO for S-ISAC), 
and that between  $V_{\rm T}(\theta)$ and $V_{\rm T-D}(\theta)$ is plotted by the blue ``+'' symbol (Impact of DISCO for T-ISAC).

From Fig.~\ref{fig5}, we can see that DISCO jamming attacks impair the sensing performance of the MUSIC algorithm.
Based on~\eqref{SensingChannel}, the DRIS-based FPJ introduces extra interference terms ${\chi}  {\boldsymbol{h}_{\rm d}^{\rm s}} \!\left({\boldsymbol{h}_{\rm D}^{\rm s}}\right)^* {\bf X}$, 
${\chi}  {\boldsymbol{h}_{\rm D}^{\rm s}} \!\left({\boldsymbol{h}_{\rm d}^{\rm s}}\right)^* {\bf X}$, and 
${\chi}  {\boldsymbol{h}_{\rm D}^{\rm s}} \!\left({\boldsymbol{h}_{\rm D}^{\rm s}}\right)^* {\bf X}$ into the received echo signals 
compared to traditional ISAC systems~\cite{FLiuISAC,FLiuISAC1}.
Since ${\boldsymbol{h}_{\rm D}^{\rm s}}$ is random and time-varying and cannot be accessed by the legitimate ISAC BS,
the sensing performance is then degraded by the DISCO jamming attacks.
From Fig.~\ref{fig5}, the resulting impact on the sensing performance also increases with the number of DRIS reflective elements. 

\section{Conclusions}\label{Conclu}
In this letter, we have investigated the trade-off between the S\&C performance in ISAC systems under DISCO jamming attacks.
The ISAC waveform design was formulated as a Pareto optimization problem with a trade-off factor. 
The CSI during the channel coherence time is no longer fixed due to the ACA introduced by the DRIS-based PFJ.
We quantified the impact of DISCO jamming attacks on the sum rate for any given ISAC waveform.
In addition, the DISCO jamming leads to biased DoA estimation, which in turn degrades the sensing performance of the system.
The amount of the bias can be increased by increasing the number of DRIS reflective elements.
To characterize the sensing performance degradation due to the DRIS-induced biases, a theoretical analysis based on the CRLB will be conducted in our future work.

\end{document}